\documentclass[structabstract]{aa}

\usepackage{graphicx}
\usepackage{natbib}
\usepackage{amsmath}
\usepackage{amssymb}
\usepackage{fancyhdr}
\usepackage{latexsym}
\usepackage{color}

\begin{document}
   \title{Centre-to-limb properties of small, photospheric quiet Sun jets}
   \titlerunning{CTL of quiet Sun jets}
   \author{F.~Rubio~da~Costa
          \inst{1} \thanks{Now at Department of Physics, Stanford University, Stanford, Ca 94305, USA}
          \and
          S. K. Solanki
          \inst{1,2}
          \and	  
          S. Danilovic
          \inst{1}
	  \and
          J. Hizberger
          \inst{1}
          \and
          V. Mart\'{i}nez-Pillet
          \inst{3}}

   \institute{Max-Planck-Institut f\"{u}r Sonnensystemforschung, Justus-von-Liebig-Weg 3, 37077 G\"{o}ttingen, Germany.\\
   		\email{frubio@stanford.edu}
         \and
              School of Space Research, Kyung Hee University,Yongin, Gyeonggi-Do,446-701, Korea.
         \and
              National Solar Observatory (NSO), Sunspot, NM 88349, USA.\\
}

   \date{Received; accepted }

  \abstract
  {Strongly Doppler-shifted Stokes $V$ profiles have been detected in the quiet Sun with the IMaX instrument on-board the SUNRISE stratospheric balloon-borne telescope. High velocities are required in order to produce such signals, hence these events have been interpreted as jets, although other sources are also possible.}
  {We aim to characterize the variation of the main properties of these events (occurrence rate, lifetime, size and velocities) with their position on the solar disk between disk centre and the solar limb.}
  {These events have been identified in Sunrise/IMaX data according to the same objective criteria at all available positions on the solar disk. Their properties were determined using standard techniques.}
  {Our study yielded a number of new insights into this phenomenon. Most importantly, the number density of these events is independent of the heliocentric angle, i.e. the investigated supersonic flows are nearly isotropically distributed.  Size and lifetime are also nearly independent of the heliocentric angle, while their intensity contrast increases towards the solar limb. The Stokes $V$ jets are associated with upflow velocities deduced from Stokes $I$, which are stronger towards the limb. Their intensity decreases with time, while their line-of-sight (LOS) velocity does not display a clear temporal evolution. Their association with linear polarization signals decreases towards the limb.}
  {The density of events appears to be independent of heliocentric angle, establishing that they are directed nearly randomly. If these events are jets triggered by magnetic reconnection between emerging magnetic flux and the ambient field, then our results suggest that there is no preferred geometry for the reconnection process.}
    
  \keywords{Sun: activity -- Sun: magnetic fields -- Sun: photosphere -- Sun: granulation}

  \maketitle

\section{Introduction}
The improvement of the spatial resolution of solar observations over the last decades has led to the detection of a number of exciting small-scale features. Observations made in the absence of seeing effects have been particularly valuable. Thus the Hinode mission \citep{2007SoPh..243....3K, 2008SoPh..249..167T}, with its high spatial resolution (0.3$''$) full Stokes vector data, has contributed very significantly to the study of the magnetism of the quiet Sun. The SUNRISE mission, during its first science flight - \citep[see][]{2011SoPh..268....1B, 2010ApJ...723L.127S}, has allowed us to probe the quiet Sun in even greater detail with seeing-free data, thanks to its even higher resolution of 0.15$''$.

It allowed, among many other results, kG magnetic fields to be resolved \citep{2010ApJ...723L.164L}; the brightness contrasts of small magnetic elements to be measured in the UV \citep{2010ApJ...723L.169R}, which is of importance for solar irradiance variations \citep{2013ARA&A..51..311S}; the dynamics of horizontal internetwork fields to be followed  \citep{2010ApJ...723L.149D} and the internal structure of network magnetic elements to be determined \citep{2012ApJ...758L..40M}. A surprising discovery was made by \citet{2010ApJ...723L.144B}, who found strong circular polarization signals at a wavelength sampling the continuum,  which is most naturally interpreted in terms of large upflow or downflow velocities, the former velocities produced by the Doppler shift of the neighbouring \ion{Fe}{I} 5250.6~\AA\ into the continuum wavelength. Such signals are short-lived, typically lasting less than 2 minutes. \citet{2011A&A...530A.111M} compared the SUNRISE results with observations from Hinode/SP, finding highly asymmetric and shifted Stokes $V$ profiles. They found that 72 \% of the features were found at the centre or edge of a granule.

A possible scenario could be that the strong line shifts are produced by magnetic reconnection between a pre-existing field in the intergranular region and emerging field \citep{2013A&A...558A..30Q}. The 3D MHD simulations of \citet{2014A&A...D_accepted} confirm that the emergence of magnetic flux and its interaction with existing field is likely to qualitatively reproduce the observational signatures. In some of the cases these are indeed due to reconnection, but many of the supersonic flows (downflows) are due to convective collapse \citep[e.g.,][]{1978ApJ...221..368P, 2008ApJ...677L.145N, 2010A&A...509A..76D}. Also, \citet{2010ApJ...723L.144B} believe that the circular polarization signal of a minority (around 20\%) of the supersonic magnetic flows that appear in evolving granules could be associated with an exploding granule \citep{1995ApJ...443..863R}.

In this paper, we characterize the centre-to-limb variations (CLV) of the main properties of these features visible in the continuum wavelegth filter position of the Stokes $V$ signal. For brevity and for consistency with the previous work \citep{2010ApJ...723L.144B}, hereafter we call them jets, although their actual physical interpretation may be more diverse. Such an investigation may allow us to set constraints on the geometry of the magnetic configuration. We use Stokes parameters acquired by the IMaX instrument onboard SUNRISE during its 2009 flight at different distances from the solar limb.

\section{Observations and identification of jets}\label{Sect:observations}
\subsection{Observations}
The SUNRISE balloon-borne solar observatory was launched on its first science flight on 8$^{th}$ June 2009 \citep{2010ApJ...723L.127S, 2011SoPh..268....1B, 2011SoPh..268..103B} carrying two instruments: SuFI (Sunrise Filter Imager, \citet{2011SoPh..268...35G}), a UV filter imager, and IMaX (Imaging Magnetograph eXperiment, \citet{2011SoPh..268...57M}), an imaging vector polarimeter operating in \ion{Fe}{I} 5250.2~\AA. IMaX has a spectral resolution of 85~m\AA\ and the effective FOV after phase diversity reconstruction is 45 $\times$ 45 arcsec$^2$. The images have a plate scale of 0.055~arcsec pixel$^{-1}$.

For the sequences used in this study, the IMaX instrument operated in its V5-6 mode, i.e. it recorded the full Stokes vector at five offsets from the centre of the \ion{Fe}{I} line ($\lambda_0=5250.217$~\AA): $\pm$40~m\AA, $\pm$80~m\AA\ and +227~m\AA. We will refer to the +227~m\AA\ offset as the continuum wavelegth position ($\lambda_{\text{c}}$) in this paper. In this mode the cadence reached by the instrument is 33 seconds and the noise level in the reconstructed data is 3-4 $\times 10^{-3}$ of the continuum intensity, $I_{\text{c}}$, at solar disk centre thanks to the six integrations carried out at each wavelength position.

\subsection{Data selection}\label{Sect:data_selection}
We used the data sets listed in Table \ref{table_data_mu}, recorded on 10$^{th}$ and 13$^{th}$ June 2009. We selected them according to $\mu=cos~\theta$, and the quality of the time series estimated from the root-mean-square (RMS) contrast of the continuum intensity images.

The data series taken at cosines of the heliocentrinc angle $\mu ~\geq~ 0.98$ have been grouped together for this study to enhance the statistics, as have the data series taken at $\mu ~=~0.46$,~0.42 and 0.40, so that we distinguish between four mean $\mu$ values. These are given in the first column of Table \ref{table_agrupation_mu}. 

\begin{table*}[!th]
\centering
\caption{Data sets used in this study.}
\begin{tabular}{c c c c c}
\hline\hline
Day & RMS Contrast (\%) & Time Range (UT) & Time Span (Minutes) & $\mu=cos \; \theta$\\
\hline
13/06/2009 & 13.66 & 16:11:14 - 16:33:24 & 22.8 minutes & 0.99\\
13/06/2009 & 13.55 & 01:24:51 - 03:12:54 & 49.9 minutes & 0.98\\
10/06/2009 & 12.92 & 08:29:16 - 09:39:38 & 69.8 minutes & 0.92\\
13/06/2009 & 13.02 & 13:43:49 - 13:59:21 & 15.5 minutes & 0.76\\
13/06/2009 & 10.56 & 01:09:53 - 01:22:04 & 12.2 minutes & 0.46\\
13/06/2009 & 9.40 & 15:42:25 - 15:55:43 & 13.3 minutes & 0.42\\
13/06/2009 & 8.99 & 17:07:45 - 17:31:35 & 23.8 minutes & 0.40\\
\hline
\end{tabular}
\label{table_data_mu}
\vspace{0.3cm}
\end{table*}

The data at $\mu=$0.76 presents less amount of events than for the rest of the data series, therefore the statistics for this data set are lower. As a consequence, this may affect to the main properties such as lifetime, contrast or area which may differ from the rest of the data sets.

\begin{table*}[!th]
\centering
\vspace{-0.3cm}
\caption{Number of jets found at the four investigated $\mu$ positions.}
\begin{tabular}{c c c c c}
\hline\hline
$<\mu>$ & $\mu=cos \; \theta$  & Total No. & No. distinct & No. distinct events\\
 &  & of events  & events & with $T>$ 33 s\\
\hline
0.99 & 0.99, 0.98 & 984 & 347 & 106\\
0.92 & 0.92 & 814 & 331 & 109\\
0.76 & 0.76 & 260 & 69 & 34\\
0.42 & 0.45, 0.42, 0.40 & 555 & 199 & 75\\
\hline
\end{tabular}
\label{table_agrupation_mu}
\end{table*}

\citet{2010ApJ...723L.144B} used two data sets of 22.7 and 31.6 minutes recorded on 9$^{th}$ June, both probably close to solar disk centre. We did not use these data series in our study because the estimation of the heliocentric angle for these data is not as precise as of the data sets used here.

To calculate with precision the pointing of the telescope taken after 10$^{th}$ June 2009 at 07:47 UT, we used the images in which the limb is visible. These were obtained as part of dedicated limb-to-limb scans with exactly this purpose in mind. We fitted the solar limb to a circle of the radius of the Sun at that time and obtained the centre of the Sun in arcsec. That allowed us to estimate the offset with respect to the imprecise values given in the header of the images. The obtained $(x,y)$ offsets were (70.9 $\pm$ 10.0, -69.3 $\pm$ 9.6) arcsec.

\subsection{Identification of the jets at Stokes $V_{\text{c}}$}\label{Sect:indentify_jets}
Following the same procedure as \citet{2010ApJ...723L.144B}, we identified all features with an unsigned Stokes $V$ signal, conservatively taken to be $ |1.25\times 10^{-2}| \; I_{\text{c}}$, corresponding to 3--4$\sigma$ at the continuum wavelength, i.e. data reconstructed in $V_{\text{c}}$ - taking into account the phase diversity calibration of the point-spread function of the optical system. In order to avoid false positives due to noise spikes, etc, we also applied a size threshold of 9 pixels, which corresponds roughly to the spatial resolution of the data. Any feature fulfilling these criteria will be called a jet from now on, for the sake of a clear nomenclature. 
Once events have been identified on individual frames, we group those which display a spatial overlap of at least one pixel on succesive frames, assigning them to the same jet. We allowed for a temporary drop in the signal level of an event, which can be caused by noise, since a signal slightly above 3$\sigma$ at one point can fall below the threshold at another time. Thus we assign features that overlap in position, but lie on frames with a temporal gap up to 66 seconds (two frames) to the same jet. Table \ref{table_agrupation_mu} lists the number of investigated jets in the four $\mu$ regimes. The third column gives the total number of events found, while the fourth column the total number of jets after $V_{\text{c}}$ features on different frames have been grouped. The fifth column refers to the number of events, already grouped, which have lifetimes $T$ longer than 33 seconds, finding that on average 66\% of events live less than 33 seconds.

The ratio between the number of individual features and jets is not affected by the position of the events on the solar disk, which suggests that the lifetimes of jets seen at different positions on the solar disk is roughly the same; see Section \ref{Sect:density_lifetime}.

Jets for which the time of appearance or disappearance is unknown were not investigated further.

\begin{figure*}[!tbh]
\centering
   \includegraphics[width=19.cm]{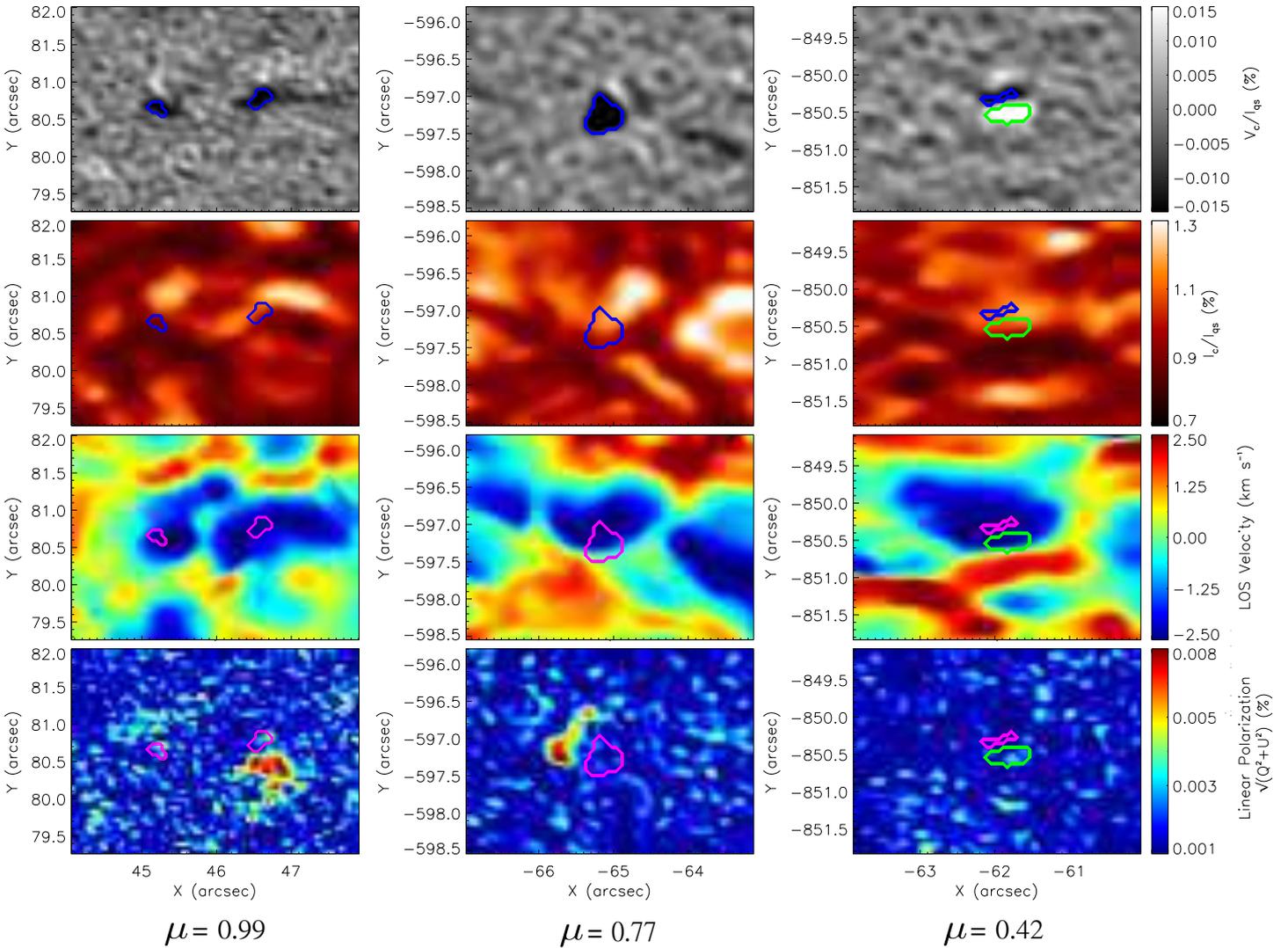}
\vspace{-0.3cm}
  \caption{Examples of jets detected in Stokes $V_{\text{c}}$ images at $\mu=0.99$ (leftmost panels), $\mu=0.76$ (middle row of panels) and $\mu=0.42$ (rightmost panels). From top to bottom the panels depict the continuum circular polarization normalized to the averaged quiet Sun continuum intensity ($V_{\text{c}}/I_{\text{qs}}$), the continuum intensity normalized to the quiet-Sun average ($I_{\text{c}}/I_{\text{qs}}$), the line-of-sight velocity and the linear polarization averaged spectrally over the \ion{Fe}{I} 5250.2 \AA\ line. The contours outline $|V_{\text{c}}| > |1.25~\times$~10$^{-2}| \; I_{\text{c}}$, where the blue contours (red in the lower two rows) signify negative values and the green contours, positive values.}
  \label{Example_identification_jets}
\end{figure*}

Fig. \ref{Example_identification_jets} illustrates jets detected at three $\mu$ values. The first row shows Stokes $V_{\text{c}}$ maps. The contours outline areas with $|V_{\text{c}}| > |1.25 \times 10^{-2}| \; I_{\text{c}}$. The other rows allow determining the continuum intensity, LOS velocity and linear polarization associated with these jets. The variation of the properties of such jets with $\mu$ will be studied in the next section.

\section{Results: Properties of the jets}\label{Section:Properties_jets}
In this Section we concentrate on the centre-to-limb variation of the main properties of the jets with lifetimes equal or longer than 66 seconds (two frames). Table \ref{table_properties} lists the number density of jets per minute and arcsec, their average lifetimes, areas and continuum contrast in the four $\mu$ bins. The listed areas, intensity contrast and LOS velocities are mean values (averaged over all events) of the maximum area reached by each event and the estimated error equals the mean standard deviation divided by the square-root of the number of jets. In the following we present the results in greater detail.

\begin{table*}[!th]
\centering
\caption{The main properties of the jets in the four $\mu$ bins.} 
\begin{tabular}{c c c c c c}
\hline\hline
$<\mu>$ & Number density of jets & Lifetime & Area at max. & Contrast at max.  & LOS Velocity\\
 & (per minute*arcsec$^2$) & & & [($I_{\text{c}}-I_{\text{qs}}$)/$I_{\text{qs}}$] & at max.\\
\hline
0.99 & (7.4 $\pm$ 0.6)*10$^{-4}$ & 86 $\pm$ 19 sec & 0.059 $\pm$ 0.019 arcsec$^2$& 0.012 $\pm$ 0.007 & -1.18 $\pm$ 0.06\\
0.92 & (8.25 $\pm$ 0.06)*10$^{-4}$ & 76 $\pm$ 16 sec & 0.059 $\pm$ 0.019 arcsec$^2$ & 0.037 $\pm$ 0.007 & -1.36 $\pm$ 0.05\\
0.76 & (7.7 $\pm$ 1.1)*10$^{-4}$ & 106 $\pm$ 25 sec & 0.056 $\pm$ 0.018 arcsec$^2$ & 0.103 $\pm$ 0.013 & -2.13 $\pm$ 0.06\\
0.42 & (5.3 $\pm$ 1.3)*10$^{-4}$ & 72 $\pm$ 16 sec & 0.052 $\pm$ 0.014 arcsec$^2$ & 0.099 $\pm$ 0.008 & -1.63 $\pm$ 0.07\\
\hline
\end{tabular}
\label{table_properties}
\end{table*}

\subsection{Morphology}
In some cases, the Stokes $V_{\text{c}}$ signal displays positive and negative values at the same time and located spatially close together. An example is shown in the right column of Fig. \ref{Example_identification_jets}. Studying a sample of twenty jets for each $\mu$ bin, we found no evidence of a trend in the ocurrence of bipolar magnetic features with $\mu$. Typically only 1-2 of the 20 jets from the sample have a bipolar $V_{\text{c}}$ signal.

The normalized continuum intensity maps ($I_{\text{c}}/I_{\text{qs}}$) in the second row of Fig. \ref{Example_identification_jets}, indicate that the jets tend to occur at the edges of granules. This trend is confirmed for all heliocentric values (e.g. Fig \ref{Example_identification_jets}) for a sample of 20 randomly chosen events in each $\mu$ bin, although a minority ($\approx$ 20\%) appear to be located in intergranular lanes. Their evolution in time exhibits that the events appear at the borders of splitting (exploding) granules.

\subsection{Number density and lifetime of the jets}\label{Sect:density_lifetime}
The average number density of events, given in Table \ref{table_properties}, is similar to the one estimated by \citet{2010ApJ...723L.144B} (7.6*10$^{-4}$ events per minute per arcsec$^2$), within a 1$\sigma$ error, irrespective of the value of $\mu$. This indicates that the jets point in all directions nearly equally, although the number of those pointing close to the horizontal direction (note that $\mu$=0.42 corresponds to a heliocentric angle $\theta$ of 65.2$^{\circ}$) may be somewhat lower.

\begin{figure}[!bh]
\centering
\vspace{-1.2cm}
\hspace{-0.7cm}
  \includegraphics[width=9.5cm]{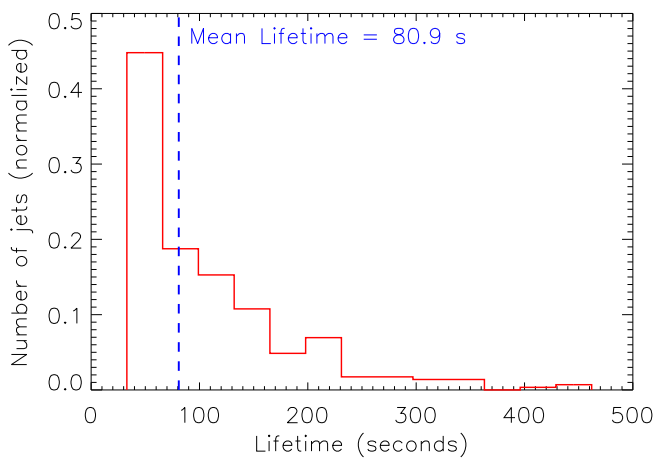}
  \caption{Lifetime histogram combining results from  all four $\mu$ bins. The density of events has been normalized to 1 and jets found on only one frame are not displayed here.}
  \label{Hist_lifetimes_norm}
\end{figure}

Lifetime distributions and average lifetimes of the jets were determined for each $\mu$ bin. Since no significant differences were discovered between the four heliocentric-angle bins, a single lifetime distribution with better statistics was calculated by combining values from all $\mu$ bins. 

Fig. \ref{Hist_lifetimes_norm} shows this combined lifetime histogram, after removing those jets seen on just a single frame. The average duration of the jets is 80.9 seconds, very similar to the value estimated by \citet{2010ApJ...723L.144B} (81.3 seconds). Assuming that the density of events follows a power law behaviour ($y=ax^b$), the power law index, $b$, has a value of -1.96. A value so close to -2 implies that the jets seen on any given snapshot are equally likely to be short-lived as long-lived. This is because although there are many more short-lived jets, the fewer longer lived ones appear on multiple frames. At a power law of -2 both groups are equally represented on a given frame.  The fact that nearly half of the jets are seen on just one frame, suggests that there are likely many jets with shorter lifetimes, which have not been detected. Roughly 9\% of the jets are visible for longer than 3 minutes.

\subsection{Area of the jets}\label{Section:Area_jets}
The averages of the jets' maximum areas decrease towards the solar limb by less than 11\%. 
Note that the decrease with $\mu$ is much slower than $\mu$, so that it is not comensurate with the foreshortening of a horizontal surface. This reduction in area with $\mu$ lies within the error bars, so that it cannot be ruled out that the jets have the same average area at all $\mu$.

\begin{figure}[!tbh]
\centering
\vspace{-1.2cm}
\hspace{-0.6cm}
  \includegraphics[width=9.5cm]{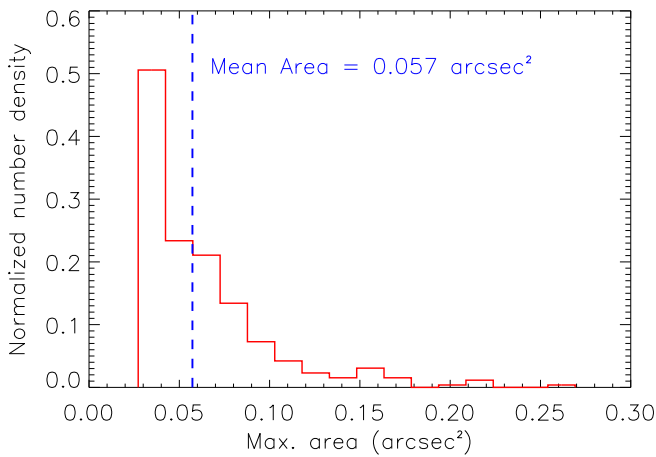}
  \caption{Histogram of maximum area reached by each jet for all studied positions on the solar disk. The area density has been normalized to 1 by dividing by the total number of jets. Only jets with areas higher than 0.027 arcsec$^2$ are displayed.}
  \label{Hist_area_norm}
\vspace{-0.3cm}
\end{figure}

The histograms of maximum projected area covered by the jets for all four $\mu$ bins are rather similar. Based on the relatively weak dependance on $\mu$, we lump all the jets together and plot a single histogram in Fig. \ref{Hist_area_norm}, which has significantly improved statistics.
The histogram of the maximum area of the jets peaks next to the detection threshold of 0.027 arcsec$^2$. This implies that there are, probably many, undetected events at smaller spatial scales. 
By fitting the distribution to a polynomial function on a logarithmic scale, the estimated power law index of -2.31 shows that there is not clear conclusive behaviour of the trend of the area of the jets since the variation of the averaged area with $\mu$ varies within the standard deviation error.
As in the previous section, having a power law of -2 would mean that small jets cover the same area on a given frame as large jets. The obtained power law of -2.31 implies that smaller jets cover a much larger area than the large jets. Furthermore, the fact that more than 43\% of the jets are smaller than 0.027 arcsec$^2$ suggests that there are likely many smaller jets which have not been detected.

This result is compatible with the findings of \citet{2014A&A...D_accepted} that most of the events retrieved from MHD simulations were on average considerably smaller than the observed events. That could also imply that a decrease in the size of the events towards the limb may be partly masked by the fact that so many events have areas close to the spatial resolution limit. 

The average of the maximum areas is about 0.057 arcsec$^2$, 1.2 times higher than the one reported by \citet{2010ApJ...723L.144B}, but it should be noted that \citet{2010ApJ...723L.144B} considered instantaneous areas of the jets and not only the maximum areas reached by the jets during their evolution.

\begin{figure}[!bh]
\centering
\vspace{-0.9cm}
\hspace{-0.6cm}
  \includegraphics[width=9.5cm]{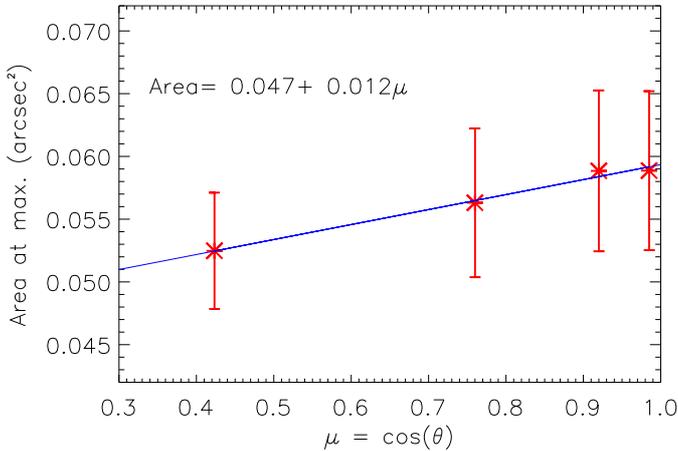}
  \caption{Centre-to-limb variation of the area of the jets (at the time in which their area is largest). Each symbol represents the average over all jets within a given $\mu$-bin. The error bars represent the standard deviations divided by the square-root of the number of jets. The solid line and the text in the figure refer to a linear regression.}
\vspace{-0.4cm}
  \label{area_jets}
\end{figure}

\subsection{Intensity contrast}\label{Sect:RMS_properties}
\begin{figure}[!tbh]
\centering
\vspace{-1.0cm}
\hspace{-0.6cm}
  \includegraphics[width=9.5cm]{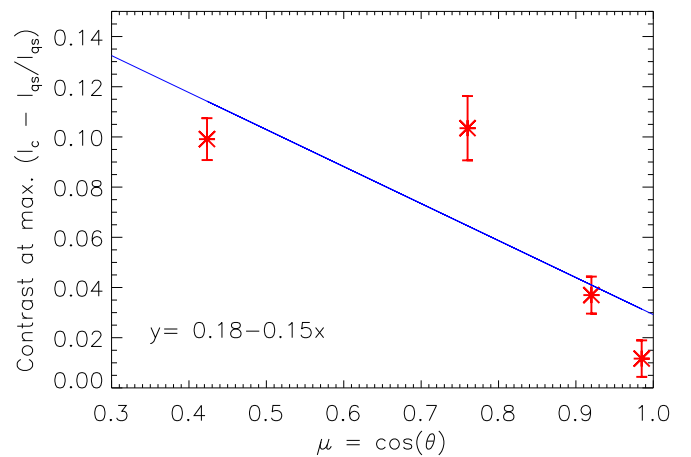}
  \caption{Same as Fig. \ref{area_jets}, but for the intensity contrast (defined as $\frac{I_{\text{c}}-I_{\text{qs}}}{I_{\text{qs}}}$) of the jets at the moment in which their surface area coverage is maximum.}
\vspace{-0.2cm}
  \label{contrast_jets_mu}
\end{figure}

The intensity contrast was defined as: $\frac{I_{\text{c}}-I_{\text{qs}}}{I_{\text{qs}}}$, where $I_{\text{qs}}$ is the averaged continuum intensity of the quiet sun image and $I_{\text{c}}$ is the continuum intensity averaged over all spatial pixels covered by the jet. Plotted in Fig. \ref{contrast_jets_mu} is the intensity contrast averaged over all jets in a certain $\mu$ bin, with the contrast for a particular jet being taken at the time step at which its surface area is largest. The contrast increases towards the limb as can be seen from Fig. \ref{contrast_jets_mu} (Table \ref{table_properties}). A similar behaviour is found if the contrasts at each time step during the lifetime of a jet are considered. 

The contrast value resulted at $<\mu>=0.76$ is well above the linear fit shown in Fig. \ref{contrast_jets_mu}. Since the represented error bars are the standard deviations divided by the square-root of the number of jet and considering that at $<\mu>=0.76$ the number of jets is lower than for the rest of the data sets, the value is within the expected behaviour.

Whereas the contrast increases towards the limb, the full width at half maximum of the distribution becomes slightly narrower, being 0.25 units at $<\mu=0.99>$ and 0.23 for at $<\mu=0.42>$. The percentage of events with intensity exceeding the averaged quiet sun intensity increases from 63\% at $<\mu>=0.99$ to 93\% at $<\mu>=0.42$.

The excess brightness of the jets is related to the fact that most of the events occur at the bright edges of the granules. The increased contrast near the limb may be related to the enhanced brightness of granule edges near the limb. Alternatively, it may be related to the origin of these events, particularly if they are associated with the convective collapse, since magnetic features tend to display larger contrasts near the limb \citep[e.g.,][]{1993SSRv...63....1S, 2004ApJ...610L.137C, 2004ApJ...607L..59K}, although this observation cannot rule out effects due to reconnection as the cause \citep{2013A&A...558A..30Q}.

We also found that jets smaller in size have higher contrast values than the bigger ones. The slope from the linear fit (-0.15 $\pm$ 0.07) is significant at the 2$\sigma$ level.

\subsection{Line-of-sight velocity} \label{Sect:LOS_vel_sect}
The line-of-sight velocity has been calculated by fitting a Gaussian profile to the four wavelength points sampling the \ion{Fe}{I} line ($\pm$40 and $\pm$80 m\AA\ relative to 5250.217 \AA) and correcting for the convection, solar rotation \citep[see][]{1984SoPh...93..219B} and shifts caused by the instrumentational configuration. The CLV of wavelegth shift introduced by the granulation was compensated by using the results of \citet{1985S&W....24..634B}. According to these, the convective blueshift decreases by 140 m/s between $\mu=0.99$ and 0.42. Note that the LOS velocities reported there are obtained from the Stokes $I$ profile.

\begin{figure}[!h]
\vspace{-1cm}
\hspace{-0.6cm}
\centering
  \includegraphics[width=9.5cm]{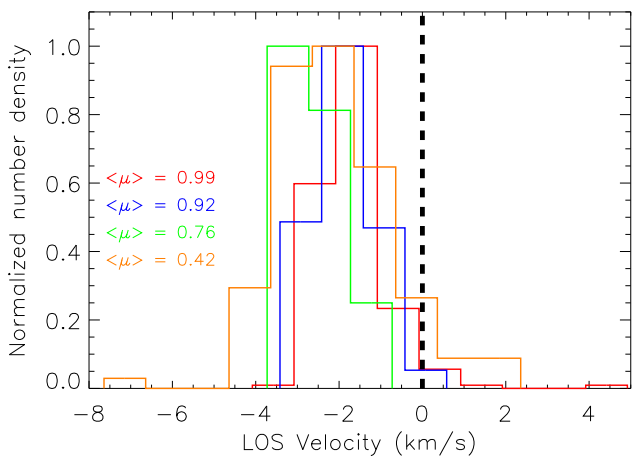}
\vspace{-0.5cm}
  \caption{LOS velocity histograms at the four analyzed $\mu$ bins. The maximum number density of events has been normalized to 1. Zero LOS velocity is indicated by the vertical dashed line.}
  \label{Hist_velocity}
\vspace{-0.3cm}
\end{figure}

Considering the jets at the moment in which their area is maximum, we found that the jets are associated with strong upflow velocities (negative values according to our reference system). This is one of the most universal characteristics, as can be judged from Fig. \ref{Hist_velocity} (cf. third row of Fig. \ref{Example_identification_jets}), which depicts histograms of the LOS velocity of the jets at the time at which their area is maximum\footnote{The results do not change when taking into account all the data of the jets at each time step.}. The upflows velocities are somewhat stronger closer to the solar limb, but this difference is not very significant. Whereas 94.1\% of the jets are associated with upflows around disk centre, this fraction has fallen to 87.7\% at $\mu=0.42$.

Since the LOS velocity is the combination of a vertical and a horizontal component, having stronger velocities for lower heliocentric angles could imply that the horizontal component is somewhat stronger or that the events seen closer to the limb are associated with stronger flows. This could be the case because close to the limb events are seen higher in the atmosphere.

Although more than 90\% of the jets with positive contrast are associated with Stokes $I$ upflows, there is no clear relationship between the continuum intensity and the LOS velocity associated with the jets (correlation coefficient $R\approx 0.3$).

\subsection{Stokes $V$ signal}
\citet{2010ApJ...723L.144B} found that near disk centre most jets are not associated with a significant unshifted Stokes $V$ signal at the event's location. To test whether this is true for other $\mu$ values, we determined $\displaystyle \sum_{i=1}^{4} \frac{1}{n}\left| \displaystyle \sum_{j=1}^{n} V_{i,j}\right|$ for each jet at the time that it covered the largest number of spatial pixels, n.

Here $i$ refers to the four wavelegth positions in the \ion{Fe}{I} 5250.217 \AA\ line and $j$ runs over the spatial pixels covered by the jet. 

9.6\% of the jets are associated to a magnetic signal $>$ 3$\sigma$ in \ion{Fe}{I} 5250.2 \AA, with no clear variation between the different heliocentric positions. The same process has been repeated for the lower noise IMaX level 1 data, i.e. prior to phase diversity image reconstruction, getting similar results.

\subsection{Relation with linear polarization signal}\label{Sect:LP_signal}
To study the relation with horizontal magnetic fields, e.g. emerging magnetic features, we considered the linear polarization signals associated with the jets, at different heliocentric angles. The phase-diversity reconstruction process increases the noise by a factor of 3 in the restored $Q$ and $U$ frames. Because the $Q$ and $U$ signals display steep histograms, dropping very rapidly with amplitude, to get the linear polarization signal we used the Stokes $Q$ and $U$ parameters of the non phase-diversity reconstructed data. The signal was calculated for each wavelength point along the \ion{Fe}{I} 5250.2 \AA, and averaged as follows: $LP = \Big(\sum\limits_{i=1}^{4} \sqrt{(Q_i^2+U_i^2)} \Big)$, where $i$ refers to the wavelength point.

The fourth row in Fig. \ref{Example_identification_jets} shows the linear polarization maps, where the contours represent $V_{\text{c}} > |1.25~\times$~10$^{-2}|$ $I_{\text{c}}$.

We identified all features with a signal higher than three times the noise level (LP $> 2\times 10^{-3}$  LP$_{\text{c}}$ $\approx 3\sigma$) in the linear polarization images. Only features found in LP with a threshold size greater than 9 pixels were taken into account.

Some of the features seen in LP appear close in time and space to the ones studied in $V_{\text{c}}$. LP signals within 0.55 arcsec distance of a jet measured in $V_{\text{c}}$ are assumed to be spatially related; linear polarization signals lying farther are assumed not to be related with the event. To temporally associate $V_{\text{c}}$ and LP signals, we assumed that they may be related within a temporal window of 5 frames ($\Delta$t=165 seconds) between the appearance/disappearance of one of the signals and the appearance/disappearance of the other. 

We allow such a large difference in time between the appearance/disappearance of $V_{\text{c}}$ and LP signals, because scenarios with a lapse between these two signals are quite easy to imagine. E.g., fresh flux appears at the solar surface and slowly rises (first appearance of LP) until it reconnects with already present field (later appearance of jet). 

\begin{figure}[!h]
\hspace{-0.48cm}
\centering
  \includegraphics[width=9.47cm]{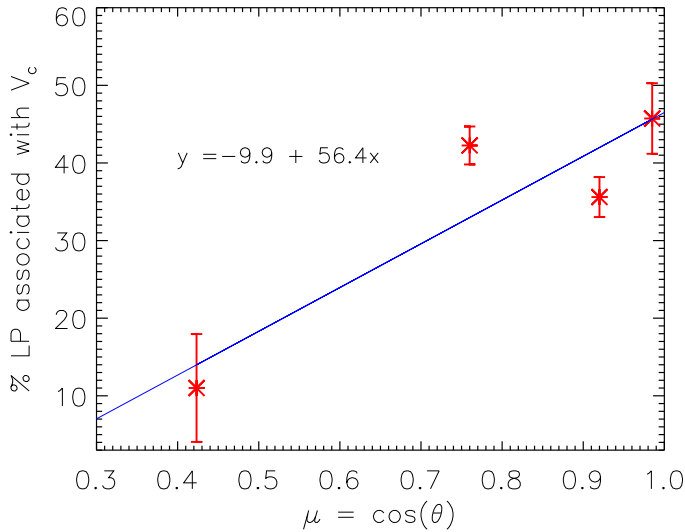}
\vspace{-0.6cm}
  \caption{Percentage of jets associated (spatially and temporally) to a linear polarization signal versus $\mu$. The error bars have been calculated taking into account the temporal and spatial uncertainties.}
  \label{density_lp_v_mu}
\vspace{-0.5cm}
\end{figure}

According to Fig. \ref{density_lp_v_mu}, at the centre of the solar disk almost half of the jets are associated with a linear polarization signal. The ratio decreases towards the solar limb, so that at $<\mu>$=0.42 only 11\% of the $V_{\text{c}}$ features lie close to a linear polarization signal, although there is a lot of scatter in the plotted relationship. Better statistics are needed to reach a reliable conclusion on this point.

More than 60\% of the LP signals associated with a jet appear before the circular polarization signal for $\mu \approx 1$ close to the 75\% of the jets that \citet{2013A&A...558A..30Q} found using a different technique), while at $<\mu>=0.42$ less than 35\% of the jets have LP emission appearing before $V_{\text{c}}$.
For more than half of these jets (56\% at $<\mu>=0.99$ to 64\% at $<\mu>=0.42$), the LP signal continues also for some time after the dissapearance of $V_{\text{c}}$.

\subsection{Evolution of the jets}\label{Sect:int_vel_time}
We normalized the lifetime of the jets from their appearance to their disappearance. Only jets living a minimum of three frames were studied at times: t=0.00 (birth), 0.50 and 1.00 (death), using spline interpolation for the cases in which the lifetime of the jets is equal to or longer than 4 frames (2 min 12 sec).

\begin{figure}[!th]
\vspace{-1.3cm}
\hspace{-0.51cm}
\centering
  \includegraphics[width=9.45cm]{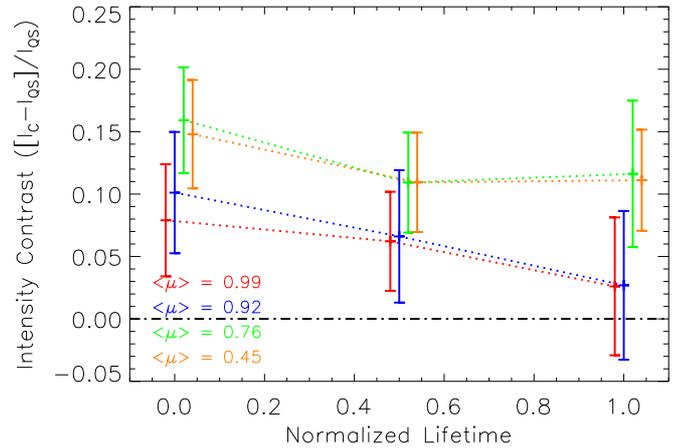}
  \caption{The averaged intensity contrast associated with the jets over their normalized lifetime for different heliocentric angles. Plotted are averages over all jets in a particular $<\mu>$ bin at a specific phase of its evolution. The values for the four $<\mu>$ have been slighly shifted on the x axis for an easier visual comparison. The lifetime is normalized  to 1. The intensity contrast $\frac{I_{\text{c}}-I_{\text{qs}}}{I_{\text{qs}}}$ has been calculated at different times [t=0 (appearance), 0.5 and 1 (disappearance)].}
\vspace{-0.6cm}
  \label{intensity_lifetime}
\end{figure}

Fig. \ref{intensity_lifetime} displays the evolution of intensity contrast (as introduced in Section \ref{Sect:RMS_properties}) over their normalized lifetime, averaged over all jets at a given $\mu$. Colors distinguish between different heliocentric angles. The error bars represent the standard deviations divided by the square-root of the number of jets. 

In all the four cases the averaged intensity contrast is largest at the time of the appearance of the jet and decreases with time, but since the uncertainty in the values is similar in magnitude to the trend, no definite conclusion can be drawn. Near the limb the intensity contrast is higher (see Section \ref{Sect:RMS_properties}) and its drop with time is not as steep as close to disk centre.

Similarly, we also considered the evolution of the LOS velocity associated with the jets. No significant trend was found, although there may be a slight tendency for the upflows to increase with time (but only below the 2$\sigma$ level).

\begin{figure}[!th]
\vspace{-1.1cm}
\hspace{-0.5cm}
\centering
     \includegraphics[width=9.45cm]{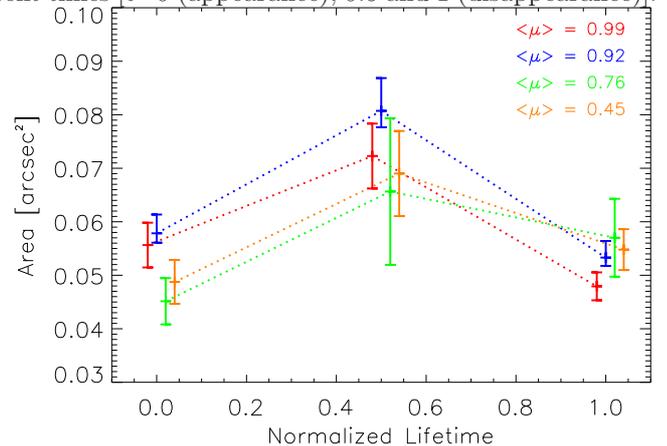}
     \caption{The averaged area of the jets over their normalized lifetime for different heliocentric angles. Plotted are averages over all jets in a particular $<\mu>$ bin at a specific phase of its evolution. The values for the four $<\mu>$ have been slighly shifted on the x axis for an easier visual comparison. The lifetime is normalized  to 1. The area has been interpolated at different times [t=0 (appearance), 0.5 and 1 (disappearance)].}
\vspace{-0.4cm}
     \label{area_lifetime}
\end{figure}

The jets' area was found to peak after roughly half of their lifetime (Fig. \ref{area_lifetime}). Jets near the limb appear to be somewhat smaller early in their life, but this difference is washed out in the course of their evolution.

\section{Conclusions}
High speed flows are present in the solar atmosphere at all heights and scales. Early observations at low resolution with Yohkoh revealed large and comparatively rare X-ray jets \citep{1994ApJ...431L..51S}. Later, signatures of bi-directional reconnection jets were detected \citep{1997SoPh..175..341I}. At the higher resolution of Hinode, smaller jets located in the quiet Sun were discovered by \citet{2007Sci...318.1591S}. Later even smaller jets were seen by the NST \citep{2010ApJ...714L..31G}. To which extent the Sunrise/IMaX high-speed events studied here and in \citet{2010ApJ...723L.144B} are similar to those detected by \citet{2010ApJ...714L..31G} and further analyzed by \citet{2011ApJ...736L..35Y} cannot be fully judged at present. However, the fact that the NST jets emerge from intergranular lanes while the `jets' we study are associated mainly with bright upflows, as well as other differences suggests that we are seeing different features.

Possibly the most important finding of this study is that the number density of jets does not vary with the heliocentric angle, meaning that the jets are equally distributed along the solar disk. This implies a close to isotropic distribution of the jet directions. Our estimated number density is similar to the one of \citet{2010ApJ...723L.144B}, within our 1$\sigma$ error.

Our lifetime values are consistent with the ones from \citet{2010ApJ...723L.144B} (80 secs) and do not change with $\mu$.

The averaged area covered by a jet decreases slightly towards the solar limb. This gentle decrease is probably the result of foreshortening. The intensity contrast increases towards the solar limb as well as does the LOS velocity obtained from the Stokes $I$ profile.

The Stokes $V_{\text{c}}$ signal is associated with a linear polarization signal. The association between horizontal and vertical magnetic fields decreases towards the solar limb. As reason for this dependence we propose the following: a horizontal magnetic field, as present at the top of a rising loop, gives a smaller linear polarization signal towards the limb with the decrease in linear polarization being proportional to $(1-\mu)^2$ for a horizontal field directed perpendicular to the nearest limb. This, together with the fact that the most common linear polarization patches lie only slighly above the noise, means that many of them drop below the noise level as their amplitude decreases (see \citet{2010ApJ...723L.149D}, for the amplitude distribution of linear polarization patches at Sunrise resolution). This can in principle explain the steep reduction of linear polarization patches associated with $V_{\text{c}}$ jets near the limb seen in Fig. \ref{density_lp_v_mu}.

\citet{2013A&A...550A.118D} simulated the reconnection between a pre-existing magnetic field and emerging fluxes resulting some brightening at the continuum of the line, associated to strong velocities and high temperatures. These results are having a similarity to the reported jets, reinforcing the assumption that are due to some quiet Sun reconnection.

The jets can also be associated with the rapid evolution of the granulation at the same location, i.e. granules that split or merge around the time the jets appear. Fast changes in granular flows can be forced by newly emerging magnetic field. This would be in agreement with the scenario proposed by \citet{2014A&A...D_accepted}. Their MHD simulations show that the observational signature of large $V_{\text{c}}$ values, i.e. the type of phenomena studied here can be produced when fresh magnetic flux emerges at small scales, mainly due to magnetic reconnection, but also due to convective collapse. However, they did not show whether the signatures are also there at different heliocentric angles.

The jets are associated mainly with upflows with velocities measured in the Stokes $I$ profile of 5250.2 \AA\ higher than 1.5 km/s on average This suggests, but does not prove, that the $V_{\text{c}}$ signal is due to a blueshift contribution of the neighbouring line \ion{Fe}{I} 5250.653 \AA: the continuum wavelength point $\lambda_c$ is situated at +227 m\AA, roughly half way between both lines. \citet{2013ApJ...768...69B} have inverted L12/2 data from Sunrise/IMaX (i.e. data at 12 wavelength points, but with higher noise and with only Stokes $V$ information). The inversions suggest that both the LOS velocity and the magnetic polarity change sign along the LOS, which the authors interpret as signatures of magnetic reconnection. 

If this interpretation is correct (they do find cases which downflows overlying upflows, which are not easily interpreted in this manner), then the reconnection jets can in principle point in an arbitrary direction, making them consistent with the near-isotropic distribution of events that we found. In addition to reconnection jets, other sources of the high-speed flows found here have been described by \citet{2014A&A...D_accepted}. They include the rapid emergence of U-loops, produced after subsurface reconnection, and the convective collapse of magnetic fields to kG values (triggered by emergence of field, but without magnetic reconnection preceding the flow). These processes are also likely to take place, but to what extent they would contribute to  the more horizontal supersonic flows still needs to be clarified. The work of \citet{2014A&A...D_accepted} further shows that most of the simulated events would remain unresolved in Sunrise/IMaX data. This points to the need for even higher resolution observations. 

\begin{acknowledgements}
The German contribution to Sunrise I was funded by the Bundesministerium f\"ur Wirtschaft und Technologie through Deutsches Zentrum f\"ur Luft- und Raumfahrt e.V. (DLR), Grant No. 50 OU 0401, and by the Innovationsfond of the President of the Max Planck Society (MPG). The Spanish contribution has been funded by the Spanish MICINN under projects ESP2006-13030-C06 and AYA2009-14105-C06 (including European FEDER funds). The HAO contribution was partly funded through NASA grant number NNX08AH38G. This work was partly supported by the BK21 plus program through the National Research Foundation (NRF) funded by the Ministry of Education of Korea.
The authors would like to express special thanks to Dr. Alex Feller for his contribution to the SUNRISE pointing correction.
\end{acknowledgements}

\bibliographystyle{aa}
\bibliography{manuscript}

\end{document}